\documentstyle[epsfig,aps]{revtex} 
\begin{document} 
\draft
\preprint{hep-th/yymmxxx} 
\title{Phenomenological equation of state and late-time cosmic acceleration} 
\author{Golam Mortuza Hossain and Parthasarathi Majumdar} 
\address{The Institute of Mathematical Sciences, Chennai 600 113,
India\\Email:golam, partha@imsc.ernet.in} 
\maketitle
\begin{abstract} 
Assuming a flat Friedmann-Robertson-Walker cosmology with a single perfect
fluid, we propose a pressure-density ratio that evolves as
a specific universal function of the scale parameter. We show that such a
ratio can indeed be consistent with several observational constraints
including those pertaining to late-time accelerated expansion. Generic
dynamical scalar field models of Dark energy (with quadratic kinetic terms
in their Lagrange density) are shown to be in accord with the proposed
equation-of-state ratio, provided the current matter density parameter
$\Omega_{m0} < 0.23$ - a value {\it not} in agreement with recent
measurements. 
\end{abstract}
\vglue .1in

The discovery of accelerated expansion of the universe from type Ia
Supernova data \cite{reis}, \cite{perl} contingent upon earlier observations
\cite{boo} indicating that the universe should be spatially flat, has led
cosmologists
to postulate a new form of energy - the so-called `Dark' energy - as the
dominant source of evolution in the present epoch. To be consistent with the
most recent observations, Dark energy is to be thought of as a perfect
barytropic cosmic fluid with an equation-of-state (pressure-density) ratio
less than -1/3. One successful candidate for this Dark energy is the
cosmological constant which can be thought of as a perfect barytropic fluid
with a linear equation of state $P_{\Lambda} = -\rho_{\Lambda}$
\cite{perl2}. 

If one considers the entire evolution of the universe, starting from
radiation-matter equilibrium through photon decoupling and matter
domination, all the way down to the present era, a single perfect barytropic
fluid with a {\it linear} equation of state $P = \kappa ~\rho$,
with a constant $\kappa$, is clearly inappropriate as a
`universal' description. The pressure-density ratio $\kappa$ changes during
this
evolution from 1/3 at the time of radiation domination, through zero
during matter domination to -1/3 or less during the current era, and
perhaps to -1 eventually, if the present acceleration is to be eternal. This
variation conflicts directly with a constant equation-of-state ratio
which implies the density evolution formula
\begin{equation}
\rho(a)~=~\rho_0~\left({ a_0 \over a} \right)^{3[1+\kappa]} ~,
\label{deno}
\end{equation}
where, $\rho_0$ is the total density in the current epoch; this is so, 
since, with a cosmological constant, the {\it lhs} of (\ref{deno}) possesses
a constant scale factor independent piece which is clearly absent in the
{\it rhs}. First order perturbations of the type
\begin{equation}
\kappa(\rho) = \kappa + \epsilon \rho~, \label{lin}
\end{equation}
where, $\epsilon$ is an infinitesimal positive
constant, lead to a cosmetic change in (\ref{deno}), given by
\begin{equation}
\rho ~=~\rho_0~\left( a_0 \over a \right)^{3(1+\kappa)} \exp - \epsilon'
\rho_0~, \label{dcor}
\end{equation}
where, ${\epsilon' \equiv {\epsilon /(1 + \kappa)}}$ is also
small, and quantities with subscript 0 indicate their values in the
current epoch. Clearly, this cannot eliminate the incompatibility discussed
above. One can at best expect the equation of state to be {\it piecewise}
linear corresponding to different epochs.  

On the other hand, it is tempting to try to model late-time cosmological
evolution in terms of a single perfect fluid; such a fluid must have a
necessarily {\it nonlinear} equation of state 
\begin{equation}
P ~=~\kappa(\rho)~\rho~ ,\label{eos}
\end{equation}
and since the energy density decays with the scale factor, the
pressure-density ratio must also evolve with the scale factor, with a
functional form chosen to satisfy the `boundary conditions'  discussed
above. This `single fluid' scenario is motivated further by the fact that
the evolution of the universe must necessarily be incremental in nature,
especially in the later epochs. It is difficult to conceive of a {\it
sudden} impulsive change in the equation of state at a specific value of
the redshift. Rather what is far more likely is a gradual transition, as the
universe expands, between
`phases' where the dominant driving source is one or other kind of
energy (including Dark energy) for a different ranges of values of the
redshift. Thus, from a classical phenomenological standpoint, it is
worthwhile to try and model cosmic evolution in late times by a single
perfect fluid with an equation-of-state ratio that changes continuously as 
a function of the scale factor. With such an evolving $\kappa$, it is not
easy to characterise what one means by `radiation domination' or `matter
domination' in the appropriate epoch. On the positive side, Dark
energy is placed on the same footing as radiation or dust matter in this
scenario. This demystification may be desirable for 
observational/phenomenological purposes.   

In this paper, we choose a specific one-parameter family
of functions of the scale factor $a$ for the pressure-density ratio; we use
several pieces of cosmological data to constrain this free parameter, and a
viable phenomenological window seems to emerge. We should mention that the
description that we have so far is a coarse-grained one wherein
details of cosmic events prior to photon decoupling have been washed
away. We find it significant though that such a simple scenario is
consistent with recent cosmological findings. We should point out that
model-independent equations of state for dark energy have indeed been
discussed recently in the literature \cite{cop}. Our proposal for the
equation of state involves not just the era of cosmic acceleration, but the
entire evolution at late-times, starting roughly with radiation-matter
decoupling.

Our choice for the functional form of the pressure-density ratio is 
\begin{equation}
\kappa(a) ~=~ \tanh\left({ a_m \over a_0} ~-~ { a \over a_0}\right) ~.
\label{ka2}
\end{equation} 
where ${a_0}$ is the present value of the scale factor and ${a_m}$ is the
only free parameter in this equation of state. This can be reexpressed in
terms of the redshift (with $a/a_0 \equiv (1+z)^{-1}$),
\begin{equation}
\kappa(z) ~=~ \tanh\left({ 1 \over {1+ z_m} } ~-~ {1 \over
{1+z}}\right)~,  
\label{red2}
\end{equation}
where ${z_m}$ is the redshift distant corresponding to the value of the
scale factor ${a_m}$. In fig. 1 below, we plot this nonlinear
equation-of-state ratio ${\kappa(z)}$ as a function of the redshift distance
${z}$ for a band of values of the parameter $z_m$.  
\begin{center}  
\epsfxsize=100mm
\epsfbox{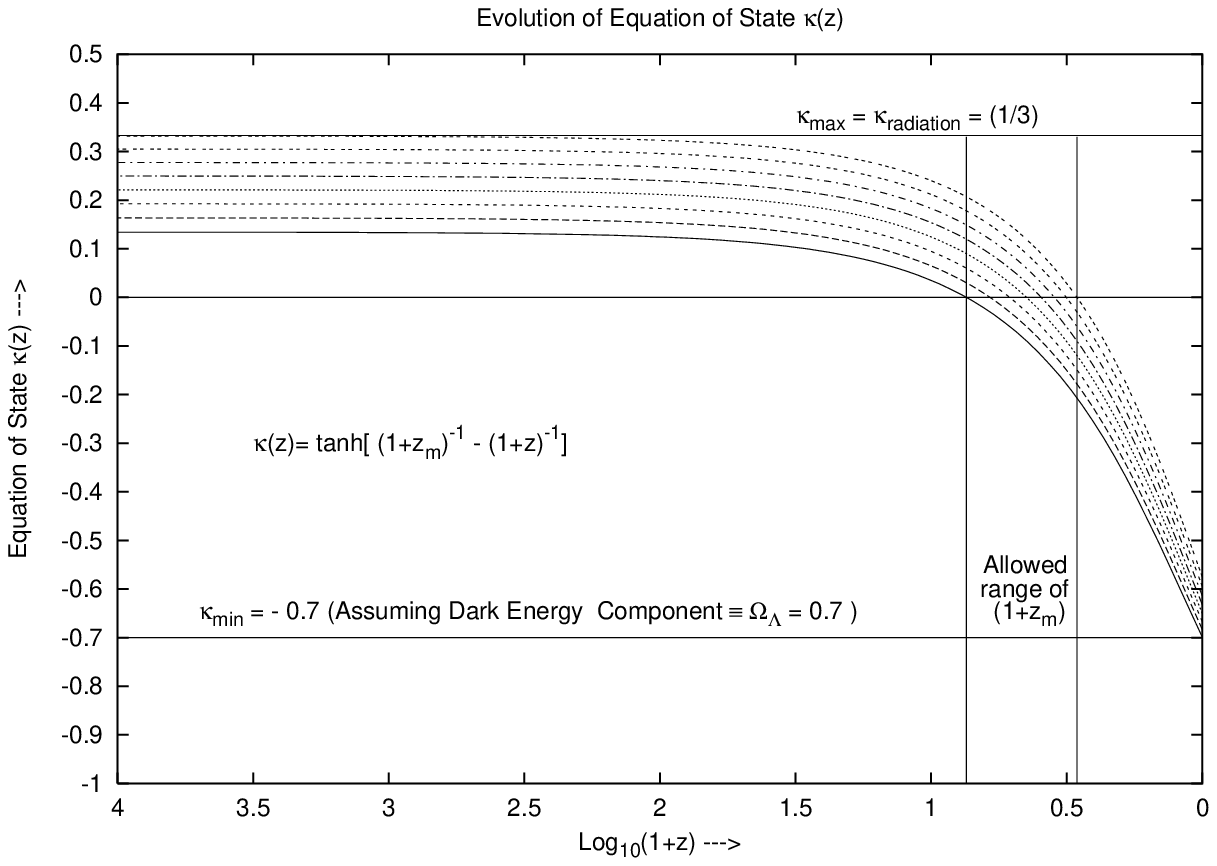}

{ Figure 1. Evolution of $\kappa(z)$ vs. $z$ for $(a_m/a_0)=0.135 ~to~
0.34$.}
\end{center}
The family of curves of the equation-of-state ratio $\kappa$ as a function
of the redshift $z$ shown in Fig. 1 corresponds to the `allowed region': the
uppermost and lowermost curves of $\kappa(z)$ correspond
respectively, as shown on the graph, to $\kappa_{max}$ depicting (partial)
radiation domination at $z=10^4$ or higher, and to $\kappa_{min}$
corresponding to the domination of Dark energy (about 70\% of the total
energy density for a flat FRW universe, according to recent observations 
\cite{perl2}) in
the present epoch (indicated by the subscript 0). The `allowed region' in
Fig. 1 restricts the parameter $a_m/a_0$ (or alternatively, $z_m$) in our
expression (\ref{ka2}) for the equation-of-state ratio to the range
$[0.175, 0.34]$. It remains to be seen whether observed values of other
cosmological parameters constrain this range even further.

For brevity of expression, we rescale $a$ by $a_0$ in the sequel. In terms
of the rescaled $a$, the overall density $\rho(a)$ evolves, via
the Friedmann equations appropriate to a flat FRW universe, according to
\begin{equation}
{{\rho(a)} \over \rho_0}~=~\left[a^{-1} \exp \int^1_a
{{d{\hat a}} \over {\hat a}}~ \kappa({\hat a})\right]^3 ~ ,
\label{dena}
\end{equation}
where $\kappa(a)$ is given, in its turn, by (\ref{ka2}). The first factor in
(\ref{dena}) exhibits the evolution of the standard
density parameter for pressureless dust matter  in absence of any Dark
energy. Thus, our proposal (\ref{ka2}) of the scale factor dependent
equation-of-state ratio  $\kappa(a)$ leads to the additional correction
factor which incorporates the effect of Dark energy. We have of course
assumed here that the total density in the current epoch is precisely equal
to the critical density. Eq. (\ref{dena}) will be useful in examining
possible constraints on the free parameter $a_m$ from cosmological data.

One further constraint is the product $H_0 t_0$ of the current value of the
Hubble parameter and the age of the universe \cite{chab}. The Friedmann
equations with a
flat FRW ansatz leads to the relation
\begin{equation}
H_0~t_0~=~\int^1_0~{da \over a}~\left( {\rho(a)\over \rho_0} 
\right)^{-1/2}~, \label{age}
\end{equation}
where, $\rho(a)$ is given by (\ref{dena}) and, as earlier,
$\kappa(a)$ is given by (\ref{ka2}). The {\it lhs} of (\ref{age}) is
known to lie in the range $0.93 \pm 0.20$. according to measurements
reported in \cite{chab}. The nested integrals on the {\it rhs} of
(\ref{age}) can be performed numerically to extract an `allowed range' of
values for the free parameter
$a_m/a_0$. Unfortunately, it turns out that this range is actually {\it
larger} than the allowed range extracted above from partial radiation
domination at redshifts larger than 10,000 and from the current observed
value of 70\%  of the Dark energy density parameter for a flat
universe. This range of allowed values is plotted in Fig. 2 below, alongside
constraints from other sources, as we now proceed to explain.

Perhaps the strongest constraint on $a_m$ comes from the angular position of
the first acoustic (`Doppler') peak in the angular distribution of the 
Cosmic
Microwave Background  Radiation (CMBR) temperature. Anisotropic temperature
fluctuations arise due to fluctuations in the gravitational potential
due to density fluctuations in the cosmic fluid. The latter, in their
turn, produce pressure fluctuations that propagate as sound waves. As
the universe expands, pressure fluctuations decrease due to changes
in the nature of the cosmic fluid, leading to attenuation of sound
waves. The proper distance travelled by such waves signifies a sonic
horizon, whose existence is inferred through the first (and largest) peak in
the angular
distribution spectrum of CMBR. Such a peak has already been
observed in the anisotropic temperature fluctuations of the CMBR by several
groups starting with COBE \cite{cobe} through BOOMERANG and MAXIMA 
\cite{boo}, \cite{maxi}, \cite{map}. Our task here is to compute the angular
position of
the first acoustic peak within the single fluid picture with our proposed
equation-of-state ratio (\ref{ka2}), as a function of the free parameter
$a_m$. Comparison with recent data on this acoustic peak is then used to
constrain $a_m$. The reason that this constraint is the strongest is linked
to the fact that the CMBR temperature anisotropy has been (and is
being) probed to an accuracy of better than $10^{-5}$. 

The location of the first acoustic peak is expressed in terms of the angle
$\theta$ subtended by the {\it sonic horizon}. This is defined
as \cite{sw1}, \cite{fram}
\begin{equation}
\theta ~\equiv~ {d_{sh} \over d_A} ~, \label{aco}
\end{equation}
where, $d_{sh}$ is the radius of the sonic horizon, defined as
the proper distance traversed by sound waves in the fluid medium
until they are attenuated due to reduction in pressure fluctuations, and
$d_A$ is the proper angular diameter distance of the sonic horizon. $d_{sh}$ 
is given by $d_{sh}=a_s~r_{sh}$ where $a_s$ is the scale
factor at which sound stops propagating: $c_s(a_s)=0$.  This can be computed
from the coordinate size $r_{sh}$ given by 
\begin{equation}
r_{sh}~=~\int_{a_s}^1 da~ {c_s \over {\dot a}}~, \label{son}
\end{equation}
where, the velocity of sound $c_s$, defined in the standard fashion as
$c_s^2 \equiv dP/d\rho$, is given, in our model by,
\begin{equation}
c_s^2(a)~=~{a \over 3} ~+~\kappa(a)~(1~-~{a \over 3})~. \label{cs}
\end{equation}
It is easy to see that $c_s(a_s)=0$ gives an analytical relation between
$a_m$ and $a_s$,
\begin{equation}
a_m~=~a_s~-~\tanh^{-1} \left[ {{a_s/3 } \over {1-a_s/3}}
\right]~. \label{sm}
\end{equation}
The angular diameter distance is given by the length of the null
geodesic traversed by photons travelling from the sonic horizon towards the
earth, scaled by the scale factor at sound attenuation: $d_A=a_s~r_p$, with
the coordinate path length $r_p$ being given
by 
\begin{equation}
r_p~=~\int_{a_s}^1~{da \over a~{\dot a}}~.\label{da}
\end{equation}
Substituting eq.s (\ref{son}) and (\ref{da}) into (\ref{aco}), we obtain
\begin{equation}
\theta (a_s) ~=~ {{\int_0^{a_s}~da~{c_s \over {\dot a}}} \over {\int_{a_s}^1~da {1
\over {a {\dot a}}}}}~. \label{the}
\end{equation}
The order $l_1$ of the multipole contributing to the first acoustic peak
is related to $\theta$ by the relation $\theta(a_s)=2\pi/l_1$. Recent
measurements of functions of the temperature distribution have constrained
$l_1$ to be in the range $212 \pm 17$. Using the formula (\ref{the}) for
$\theta(a_s)$, one can determine (numerically) the range of allowed
values of $a_s$ and hence, using (\ref{sm}), obtain the admissible range of
values of $a_m$. One obtains the range $a_m = 0.172 \pm 0.005$, which
falls well within the range allowed by the other cosmological
constraints. In fig. 2 we exhibit these allowed ranges of values for this
parameter constrained by cosmological data. 
\begin{center}  
\epsfxsize=100mm
\epsfbox{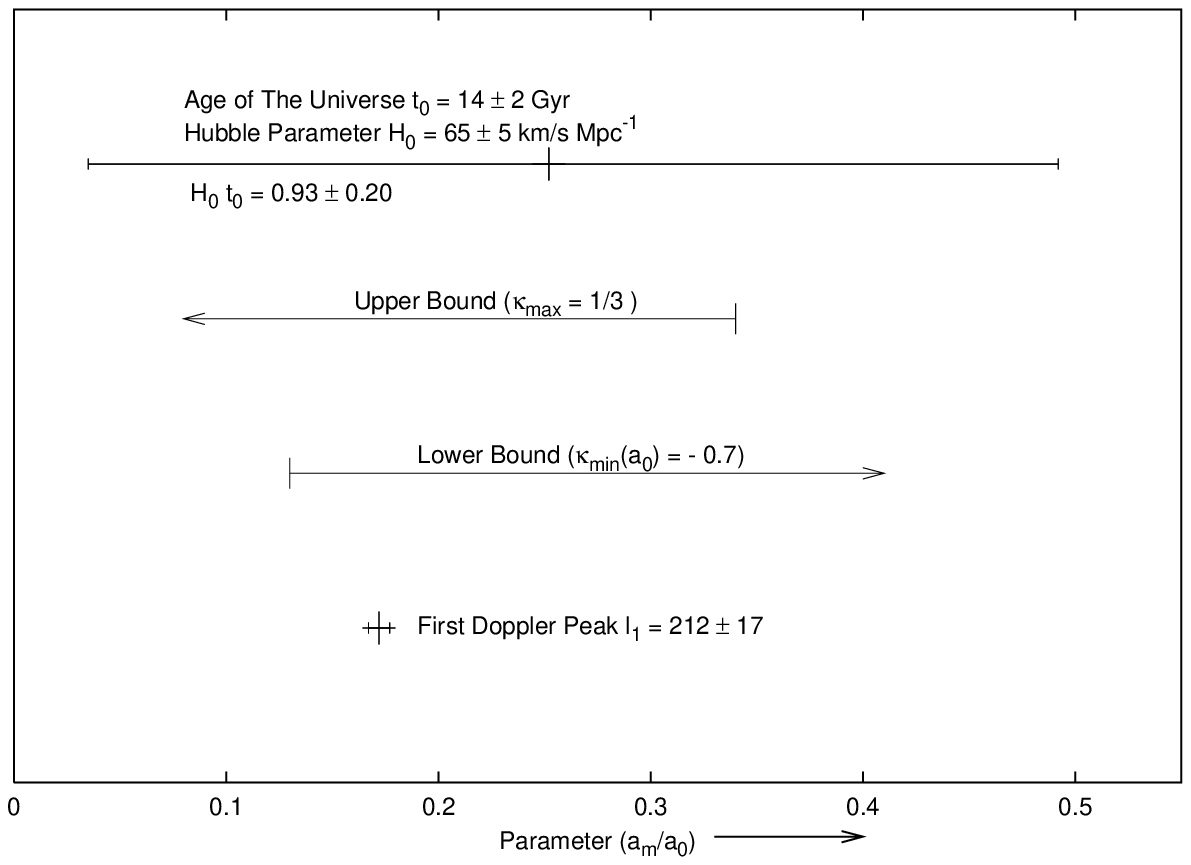}

{Figure 2. Allowed range of $a_m$ constrained by cosmological data.}
\end{center}

Having established the viability of the phenomenological equation-of-state
ratio given by (\ref{ka2}), the issue now arises whether such an equation of
state can actually be obtained from a dynamical scalar field model. In this
letter we confine our attention to models with a standard quadratic kinetic
energy term; tachyonic models now emerging as popular candidates for Dark
energy have non-standard actions with infinite-order derivative terms and
lie outside the purview of this letter. The same is the case with `tracker'
models with exotic kinetic energy structures. We shall consider these in a
future assay.  

We assume that at late times, the major contribution to the total energy
density comes from pressureless dust matter and the scalar field such that
${ \rho ~=~ \rho_m + \rho_\phi}$. Assuming a spatially homogeneous scalar
field with a potential $V(\phi)$, the density and pressure are given by 
\begin{eqnarray}
\rho_{\phi} ~&=&~\frac12 {\dot \phi}^2~+~V(\phi)~ \nonumber \\
P_{\phi} ~&=&~\frac12 {\dot \phi}^2~ -~V(\phi) ~. \label{dp}
\end{eqnarray}
This implies that
\begin{equation}
\frac{1}{2} {\dot \phi}^2 ~=~ V ~+~ \kappa(a)\rho(a)~. 
\label{ke}
\end{equation}
The scalar field obeys the field equation (in a flat FRW geometry)
\begin{equation}
\ddot\phi ~+~ 3({\dot a \over a})~{\dot \phi}~
 + ~ {dV \over d\phi} ~=~ 0~. \label{eom}
\end{equation}  
The potential $V(\phi)$ can be eliminated from the field equation using
eq.s (\ref{dp}) and (\ref{ke}), yielding a first order linear
differential equation in ${\dot \phi}^2$
\begin{equation}
{d \over dt}[a^3~{\dot \phi}^2]~=~a^3~{d \over dt}[\kappa
\rho]~. \label{ph}
\end{equation}
The solution may be written as
\begin{equation}
\frac{1}{2} {\dot \phi}^2 ~=~ \frac{1}{2}
\left[ { \rho_{D0} \over a^3 } + \kappa(a)\rho(a)
 + \frac{3}{a^3} \int_{a}^{1} dx x^2 \kappa(x)\rho(x) \right] ~, \label{ke1}
\end{equation}
where, $\rho_{D0} \equiv \rho_0-\rho_{m0}$ is the density of Dark energy
at the present epoch.

Now, in the range $[a_m~,~1]$ of the scale factor, $\kappa(a)$ is clearly
negative, so that the second and third terms on the {\it rhs} of eq.
(\ref{ke1}) produce negative contributions, while the first term
contributes positively. Since the {\it lhs} must always be positive, this
leads (numerically) to a rather stringent constraint on the Dark energy
density parameter at the present epoch: $\Omega_{D0} ~>~ 0.77$. That is to
say, {\it a generic scalar field model of Dark energy can be consistent
with the equation of state (\ref{ka2}) only for a rather large Dark energy
density parameter, or what is equivalent, a current matter density
parameter $\Omega_{m0} < 0.23$}. This is  lower than the measured
value of around $1/3$. 

It follows that, if our phenomenological equation of state is correct, a
generic dynamical scalar field model of Dark energy (like, for instance,
some Quintessence models), with quadratic kinetic energy, is unlikely to
be an appropriate description.  In other words, scalar field models may
not be able to explain why there exists a cosmological constant that is so
small when compared to the vacuum energy of the standard model, and is yet
almost the sole driving force of cosmic evolution at present. That the
proposed equation-of-state ratio is consistent with recently measured
values of several key cosmological parameters lends more credence to this
speculation.

We thank G. Date for several useful discussions. We also thank A. Ganguly
for discussions and for bringing ref. 5 to our attention.

\end{document}